\documentstyle[aps,graphicx]{revtex}  
\def\ea{{\it et al.}}

\def\BAF{BaF$_2$ }

\def\PBW{PbWO$_4$ }
\def\PBWN{PbWO$_4$}

\begin{document}        
\baselineskip 14pt
\title{PRECISION CRYSTAL CALORIMETRY IN HIGH ENERGY PHYSICS}
\author{Ren-yuan Zhu}
\address{Physics Department, California 
        Institute of Technology, Pasadena, CA 91125, U.S.A.}
\maketitle              
\begin{abstract}        
Crystal Calorimetry is widely used in high energy physics
because of its precision. Recent development in crystal technology 
identified two key issues to reach and maintain crystal precision: 
light response uniformity and calibration {\it in situ}. Crystal
radiation damage is understood. While the damage in alkali halides 
is found to be caused by the oxygen/hydroxyl contamination, it is 
the structure defects, such as oxygen vacancies, cause damage 
in oxides. 
\end{abstract}          

\section{INTRODUCTION}
\label{sec:int}

Total absorption shower counters made of inorganic scintillating
crystals have been known for decades for their superb energy 
resolution and detection efficiency. In high energy and nuclear 
physics, large arrays of scintillating crystals have been assembled 
for precision measurements of photons and electrons.
Recently, several crystal calorimeters have been designed and are 
under construction for the next generation of high energy physics 
experiment.  Table~\ref{tab:calorimeter} summarizes 
design parameters for these crystal calorimeters. One notes that each of 
these calorimeters requires several cubic meters of high quality 
crystals. 

CsI(Tl) crystals are known to have high light yield, so was chosen
by two B Factory experiments where low noise is essential for low end
of energy reach. \PBW crystals are distinguished with their high density,
short radiation length and small Moli\`{e}re radius, so was chosen by 
CMS experiment to construct a compact crystal calorimeter of 25 radiation 
length. The low light yield of \PBW crystals can be overcome by gains of the 
photo-detector, such as PMTs and avalanche photodiodes (APD).
The unique physics capability of crystal calorimetry is the result 
of its superb energy resolution, hermetic coverage and fine granularity
\cite{bib:arnps}. Recently designed crystal calorimeters, however, 
face a new challenge: radiation damage caused by increased center of mass 
energy and luminosity. While dose rate is expected to be a few rad per 
day for CsI(Tl) crystals at two B Factories, it would reach 15 to 600 rad 
per hour for \PBW crystals at LHC. 

This paper discusses two key issues related to precision 
of crystal calorimetry {\it in situ}, and cause and cure of 
radiation damage in crystals. Light response uniformity and calibration
{\it in situ} are discussed in Sections~\ref{sec:uni} and \ref{sec:cal}.
Effect of radiation damage is elaborated in  Section~\ref{sec:rad}. 
Section~\ref{sec:mechanism} discusses damage mechanism for alkali halides, 
such as \BAF and CsI, and oxides, such as bithmuth gemanade 
(Bi$_4$Ge$_3$O$_{12}$, BGO) and \PBWN. Finally, a brief summary 
is given in Section~\ref{sec:summary}.

All measurements, except specified otherwise, were 
carried out at Caltech with samples from Beijing Glass Research 
Institute (BGRI), Bogoroditsk Techno-Chemical Plant (BTCP), 
Khar'kov and Shanghai Institute of Ceramics (SIC).

\begin{table}[hbt]
\caption{Parameters of Recently Designed Crystal Calorimeters}
\label{tab:calorimeter}
\begin{tabular}{lrrrr}
Experiment          & KTeV &  $BaBar$    & BELLE&   CMS\\
Laboratory          & FNAL &  SLAC     & KEK   &  CERN\\
\tableline 
Crystal Type        & CsI  &  CsI(Tl)  &CsI(Tl)&  \PBW   \\
B-Field (T)         &  -   & 1.5     & 1.0   &   4.0   \\
Inner Radius (m)    &  -   & 1.0     & 1.25  &   1.29   \\
Number of Crystals  &3,300 &6,580   & 8,800 & 83,300  \\
Crystal Depth (X$_0$) & 27 &  16 to 17.5& 16.2  &   25    \\
Crystal Volume (m$^3$)& 2 &  5.9       & 9.5   &   11   \\
Light Output (p.e./MeV) & 40 & 5,000    & 5,000 &   2  \\
Photosensor         & PMT  &   Si PD   & Si PD & APD\tablenote{Avalanche 
photodiode.}  \\
Gain of Photosensor &4,000 &   1       &   1   &   50     \\
Noise per Channel (MeV) & small &   0.15 &  0.2    &  30      \\
Dynamic Range       &$10^4$ &$10^4$  &$10^4$ & $10^5$  \\
\end{tabular} 
\end{table}

\section{Crystal Light Response Uniformity}
\label{sec:uni}

GEANT simulation shows that an adequate light response uniformity profile 
is a key to precision of a crystal calorimeter. 

\begin{figure}[ht]
\vspace{8.5cm}
\includegraphics{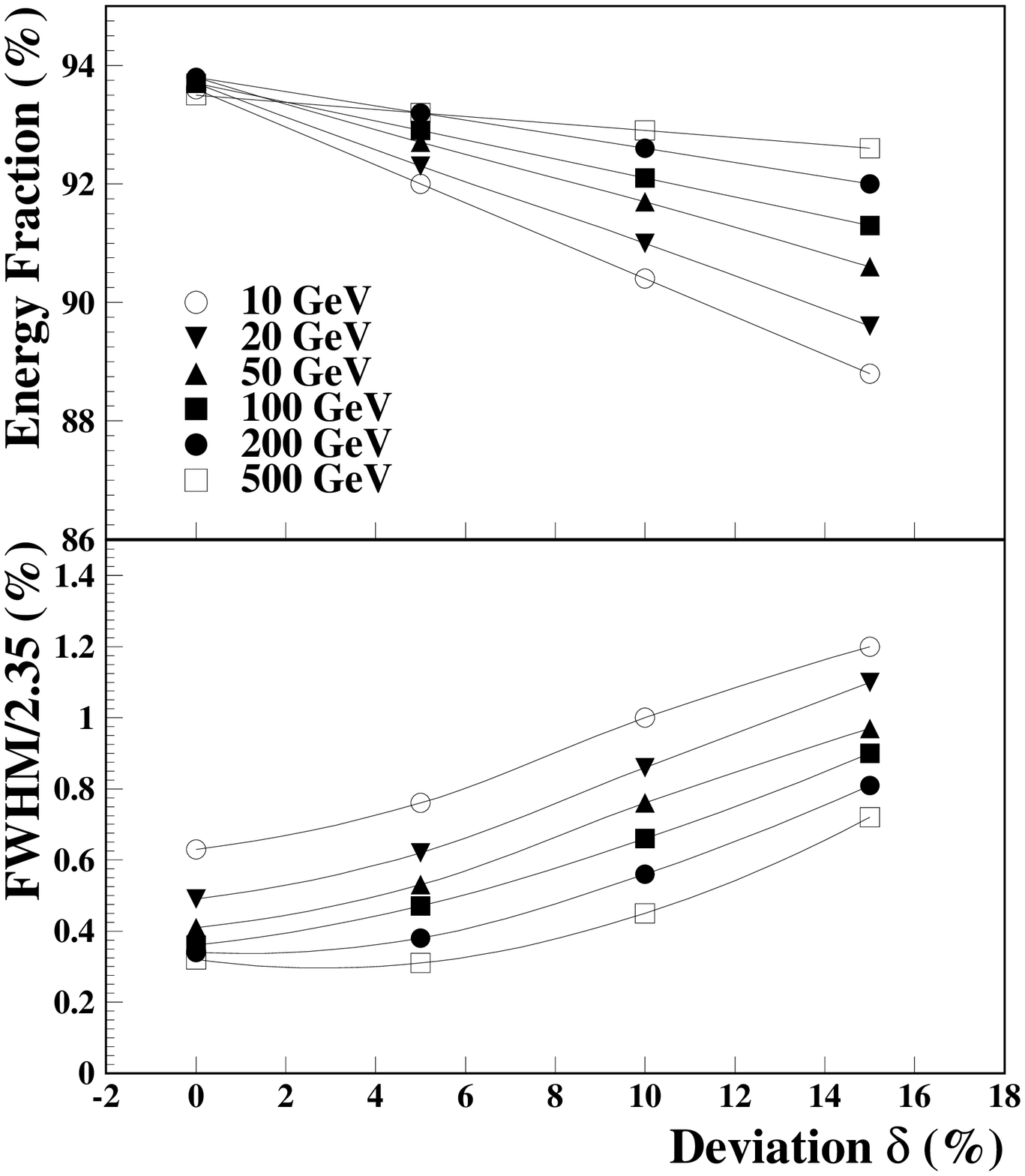}
\includegraphics{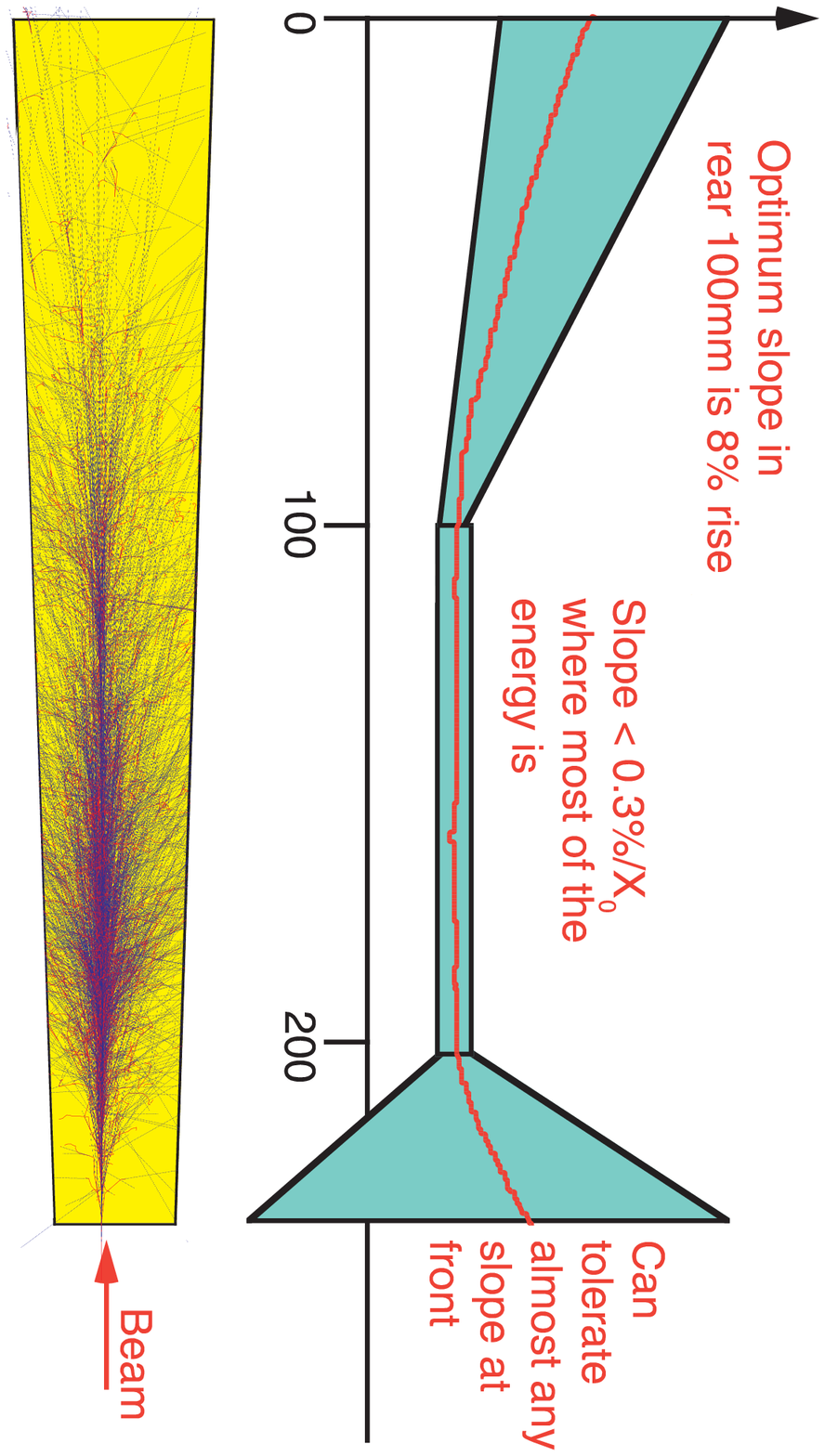}
\caption{Left: Effect of light response uniformity predicted by a
 GEANT simulation \protect\cite{bib:zhu_baf}.
 Right: Specification of light response uniformity profile for
 CMS \PBW crystals \protect\cite{bib:seez}.}
\label{fig:geantuni}
\end{figure}

The left side of Figure~\ref{fig:geantuni}~\cite{bib:zhu_baf} shows a 
GEANT prediction of energy fraction (top) and the intrinsic 
resolution (bottom) calculated by summing the energies 
deposited in a 3 $\times$ 3 sub-array, consisting of tapered \BAF crystals 
of 25 radiation length, as a function of the light response uniformity. 
In this simulation, light response (y) of the crystal was parametrized 
as a normalized linear function:
\begin{equation}
\frac{y}{y_{mid}} = 1+\delta(x/x_{mid}-1),
\label{eq:unit}
\end{equation}
where y$_{mid}$ represents light response at the middle of the crystal, 
$\delta$ represents deviation of light response uniformity, and 
x is the distance from the small (front) end of tapered crystal. 

While changes of amplitude of light output can be 
inter-calibrated, the loss of the energy resolution, caused by  
degradation of light response uniformity is not recoverable. 
To preserve crystal's intrinsic energy resolution light response 
uniformity thus must be kept within tolerance. According to above 
simulation, the $\delta$ value is required to be less than 5\% so that 
its contribution to the constant term of the energy resolution is less 
than 0.5\%.  A recent GEANT simulation for CMS \PBW crystals confirmed this 
conclusion. The right side of Figure~\ref{fig:geantuni}~\cite{bib:seez}  
shows specification of CMS \PBW uniformity profile. While the slope 
at the front 3 X$_0$ is not restricted, it must be kept within 
0.3\%/X$_0$ in the middle 10 X$_0$, and it is required to have 
a positive value of 8\% in the back 12 X$_0$, so that rear leakage 
at high energies can be compensated. 

\begin{figure}[ht]
\vspace{5.5cm}
\includegraphics{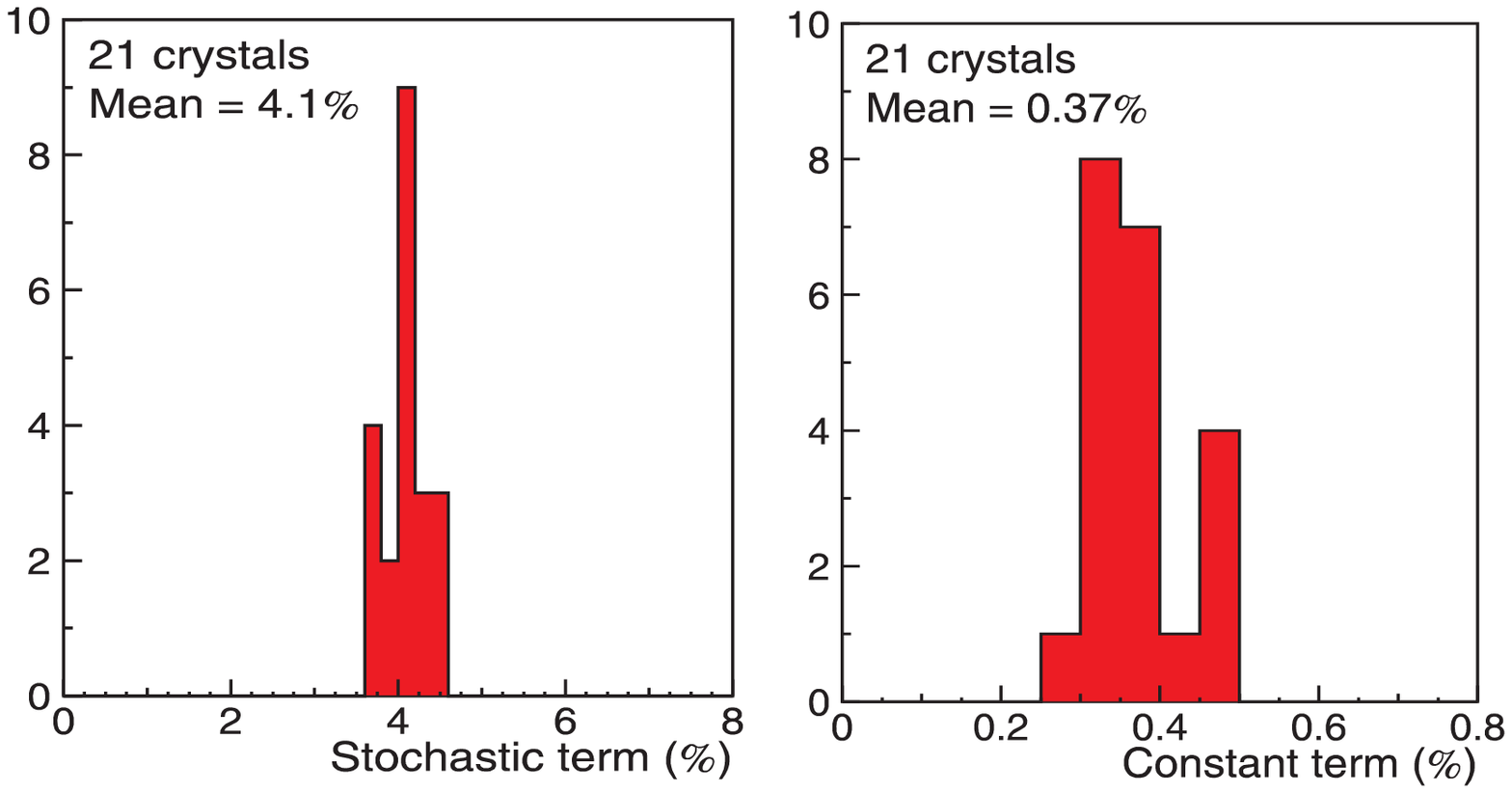}
\includegraphics{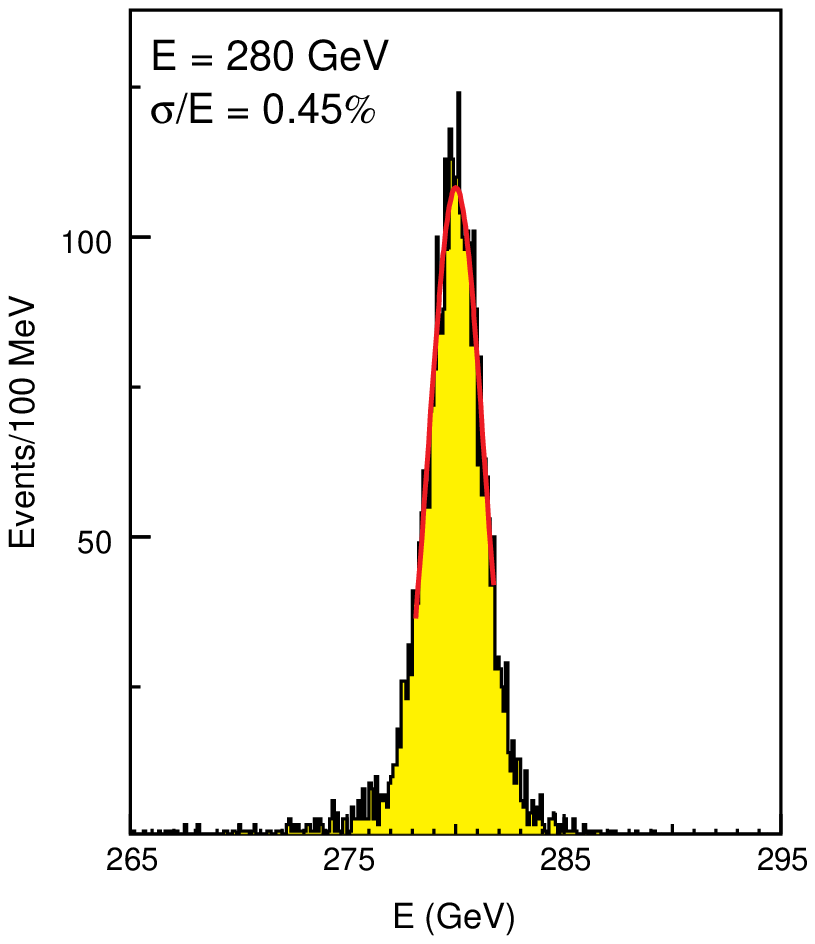}
\caption {Stochastic (Left) and constant (Middle) terms of
 energy resolution and 280 GeV electron signals (Right) obtained
 in CERN beam test.}
\label{fig:ecal_tb}
\end{figure}

By using \PBW crystals tuned according to this specification, 
an energy resolution of \( \frac{\delta E}{E} = \frac{{\bf 4.1}\%}{\sqrt{E}} 
 \oplus {\bf 0.37}\% \oplus {0.15}/E  \)
was achieved in CMS test beam at CERN using current production 
\PBW crystals with Si APD of 25 mm$^2$~\cite{bib:ecal_tb}. 
Figure~\ref{fig:ecal_tb} shows the distributions of stochastic 
(left) and constant (middle) terms of energy resolution, and 0.45\% 
energy resolution reconstructed in 3 $\times$ 3 \PBW crystals for 
280 GeV electrons (right). This 4.1\% stochastic term will be reduced 
to 3\% by using two APDs instead of one in final design~\cite{bib:ecal_tdr}. 
Note, this constant term of 0.37\% achieved does not include uncertainties 
of calibration {\it in situ}.

\ \\

\section{Precision Calibration {\it in situ}}
\label{sec:cal}

\begin{figure}[t]
\vspace{5.5cm}
\includegraphics{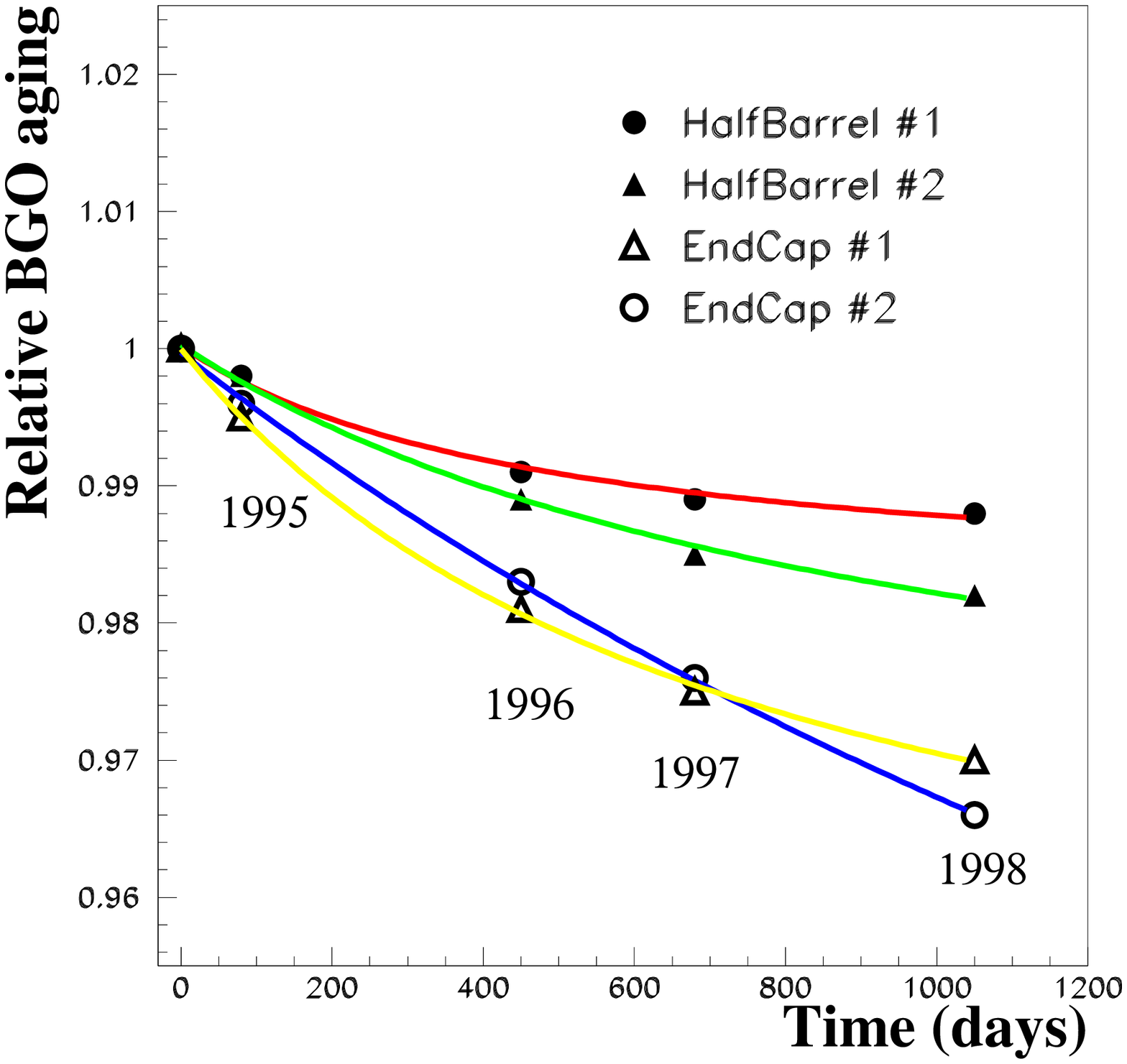}
\includegraphics{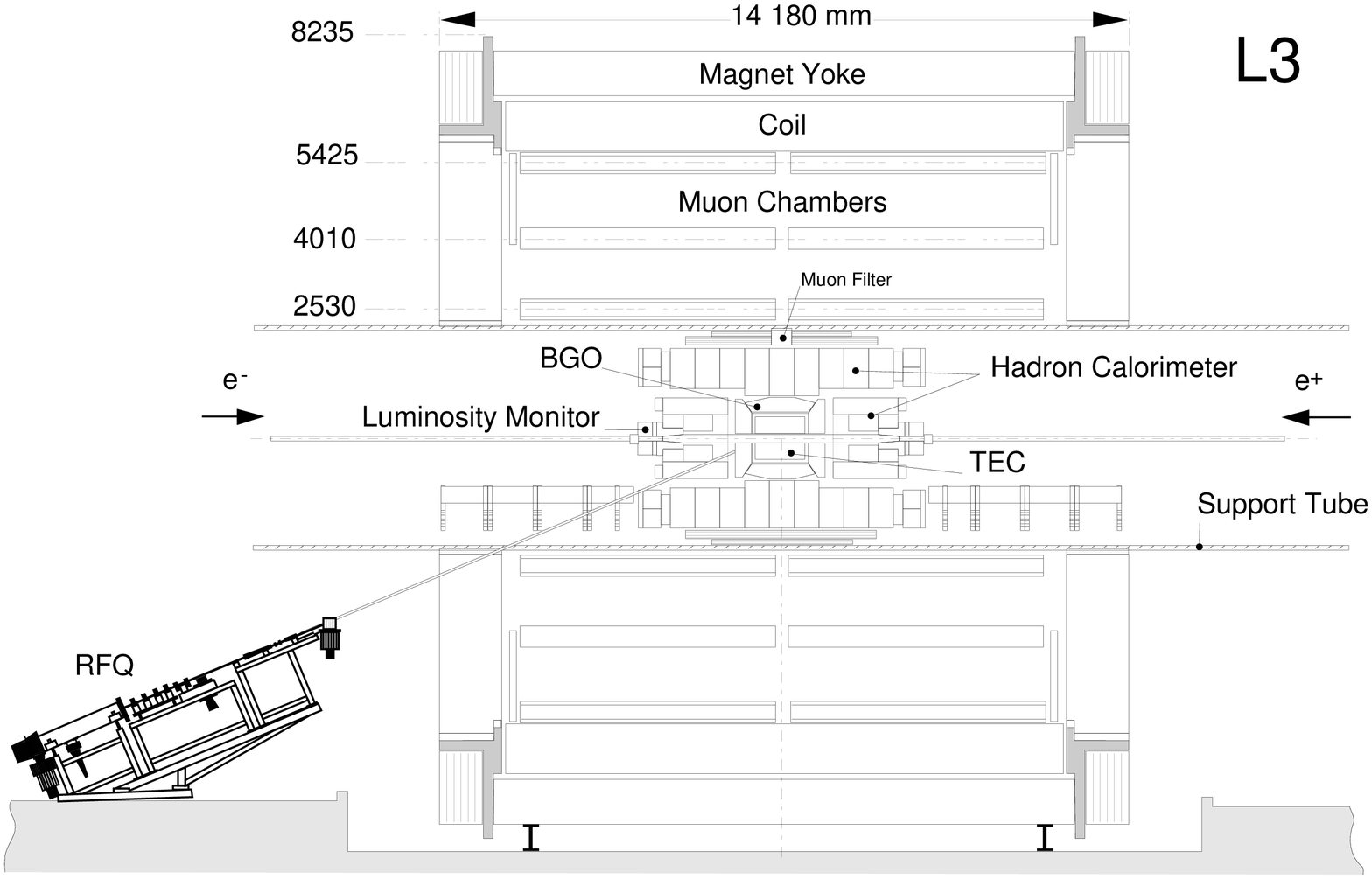}
\includegraphics{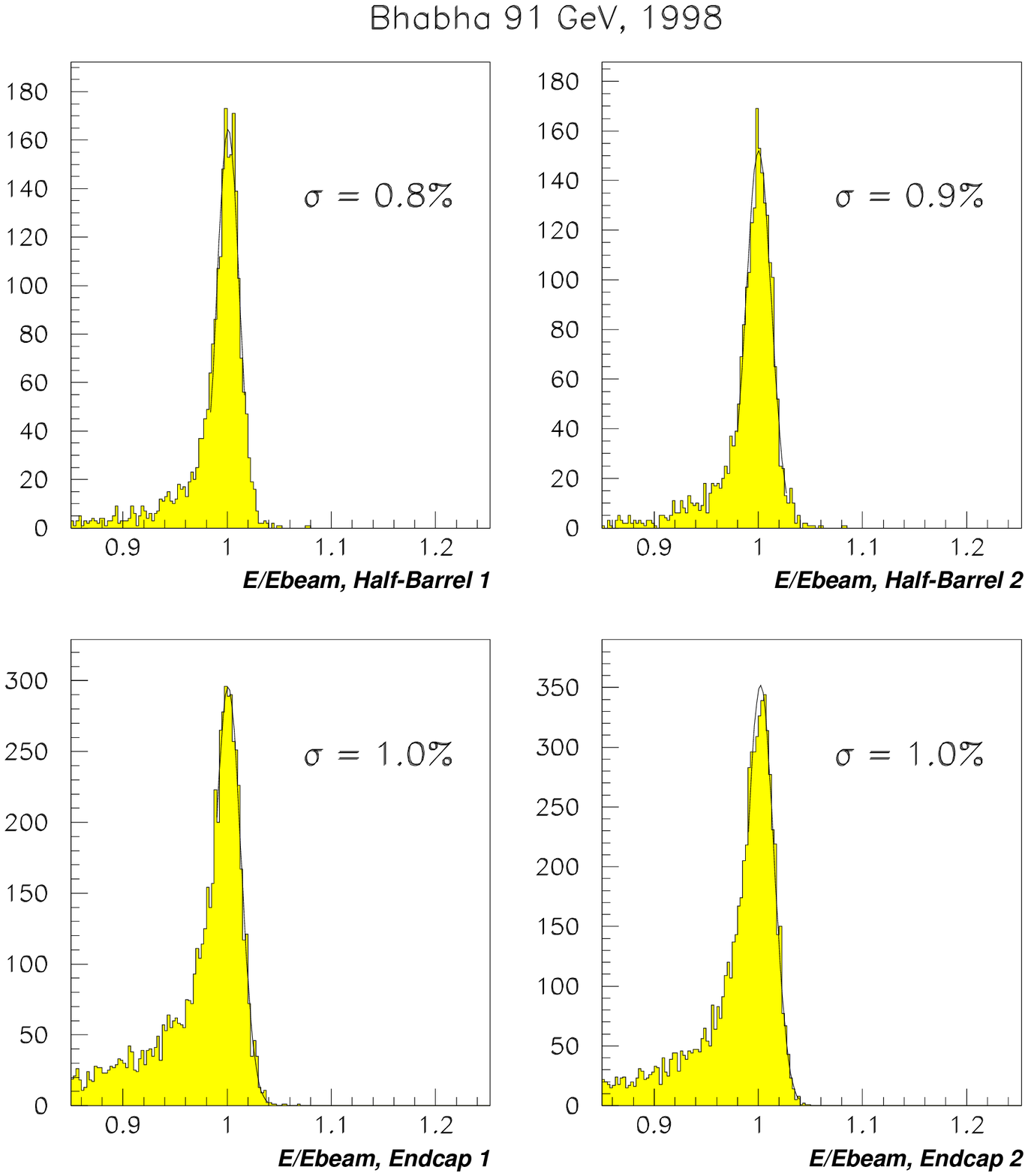}
\caption{Left: BGO aging at LEP. Middle: Layout of the L3-RFQ 
 calibration system. Right: L3 Bhabha peaks obtained with RFQ calibration.}
\label{fig:l3rfq}
\end{figure}

Precision calibration is the key factor in maintaining the 
crystal calorimetry precision {\it in situ}.
Although all individual cells of a crystal calorimeter 
may be calibrated in a test beam at several different
energies to provide a set of initial calibration constants before 
installation, the change in response over time  differs from 
one calorimeter element (crystal, photo detector, readout
chain) to the next. The left plot in Figure~\ref{fig:l3rfq}
shows BGO aging as a function of time of operation for two
half barrels and two endcaps~\cite{bib:rfq_ieee98}. 
Inter-calibrations {\it in situ} therefore are required 
to track down the evolution of each channel independently. 

Calibration {\it in situ} is most commonly achieved by using 
physics processes produced by beam, such as electrons or photons 
of known energy, electron or photon pairs reconstructible to a 
known invariant mass, E/p of electrons with momentum measured, 
and energy of minimum ionizing particles with known path length. 
Low energy $\gamma$-rays from radioactive sources or from radiative 
capture reactions are often used as a low energy calibration source. 
This is particularly important for those crystal calorimeters where 
physics processes do not occur at a high enough rate for a frequent 
calibration, such as L3 BGO \cite{bib:l3bgo}. Finally, a light pulser 
system is a useful tool to monitor the light collection in a crystal 
and the readout response. As discussed in Section~\ref{sec:rad}, 
it can also serve as inter-calibration {\it in situ}, if 
the scintillation mechanism of crystals is not damaged.

Because of limited statistics of physics processes, 
L3 experiment uses a Radiofrequency Quadrupole (RFQ)
based accelerator system for BGO crystal calibration.
The 17.6 MeV $\gamma$-rays from a radiative capture reactions
\begin{equation}
{\rm{p + ^7_3 Li \rightarrow ^8_4 Be + \gamma}} 
\label{eq:rfq_li}
\end{equation}
is used as calibration source, which was produced by
bombarding a Li target mounted inside the calorimeter 
with a proton beam. Shown in the middle of Figure~\ref{fig:l3rfq} is
the installation of RFQ calibration system in L3 detector.
Combining with Bhabha events, the RFQ system provides sub percent 
calibration {\it in situ}, as shown in the right plot of 
Figure~\ref{fig:l3rfq}~\cite{bib:rfq_ieee98}.

For recently designed crystal calorimeters listed in 
Table~\ref{tab:calorimeter}, KTeV uses E/p of electrons
from $K_L\rightarrow\pi^+e^-\nu$, $BaBar$ and BELLE will use 
electrons from Bhabha scattering,  and CMS will use E/p 
from electrons and Z $\rightarrow e^+e^-$ mass reconstruction.
In addition, $BaBar$ also uses 6.13 MeV $\gamma$-rays from
a meta stable state of $^{16}$O with $t_{1/2}$ of 7 sec, which
is produced by circulating a fluorine containing fluid through 
a neutron source, and CMS also uses a light monitoring system
to catch change of light collection in \PBW crystals {\it in situ}.
By using E/p calibration, the KTeV CsI calorimeter has achieved 
a 0.6\% resolution for electrons with energy larger than 20 GeV,
indicating an accuracy of better than 0.5\% is achieved in 
calibration {\it in situ} \cite{bib:ktevn}. The goal of CMS
experiment is to calibrate \PBW calorimeter to a similar or better
level by using physics processes combined with light monitoring.

\ \\

\section{Radiation Damage  in Scintillating Crystals}
\label{sec:rad}

All known crystal scintillators suffer from radiation damage. 
The most common damage phenomenon is the appearance of radiation-induced 
absorption bands caused by color center formation. Absorption bands 
reduce crystal's light attenuation length (LAL), and hence the light output. 
Color center formation, however, may or may not cause a degradation
of the light response uniformity. Radiation also causes phosphorescence 
(afterglow), which leads to an increase of readout noise. 
Additional effect may include a reduced intrinsic scintillation light yield 
(damage of scintillation mechanism), which would lead to a reduced 
light output and a deformation of the light response uniformity.
Damage may recover under room temperature, which leads to a
so called ``dose rate dependence''. Finally, thermal annealing and optical 
bleaching may be effective in eliminating color centers in crystals. 
Reference~\cite{bib:zhu_nim} and the references therein
provide detailed information for readers with interest. 
Because of limited scope we only highlight two points 
below.

\begin{figure}[ht]
\vspace{8.cm}
\includegraphics{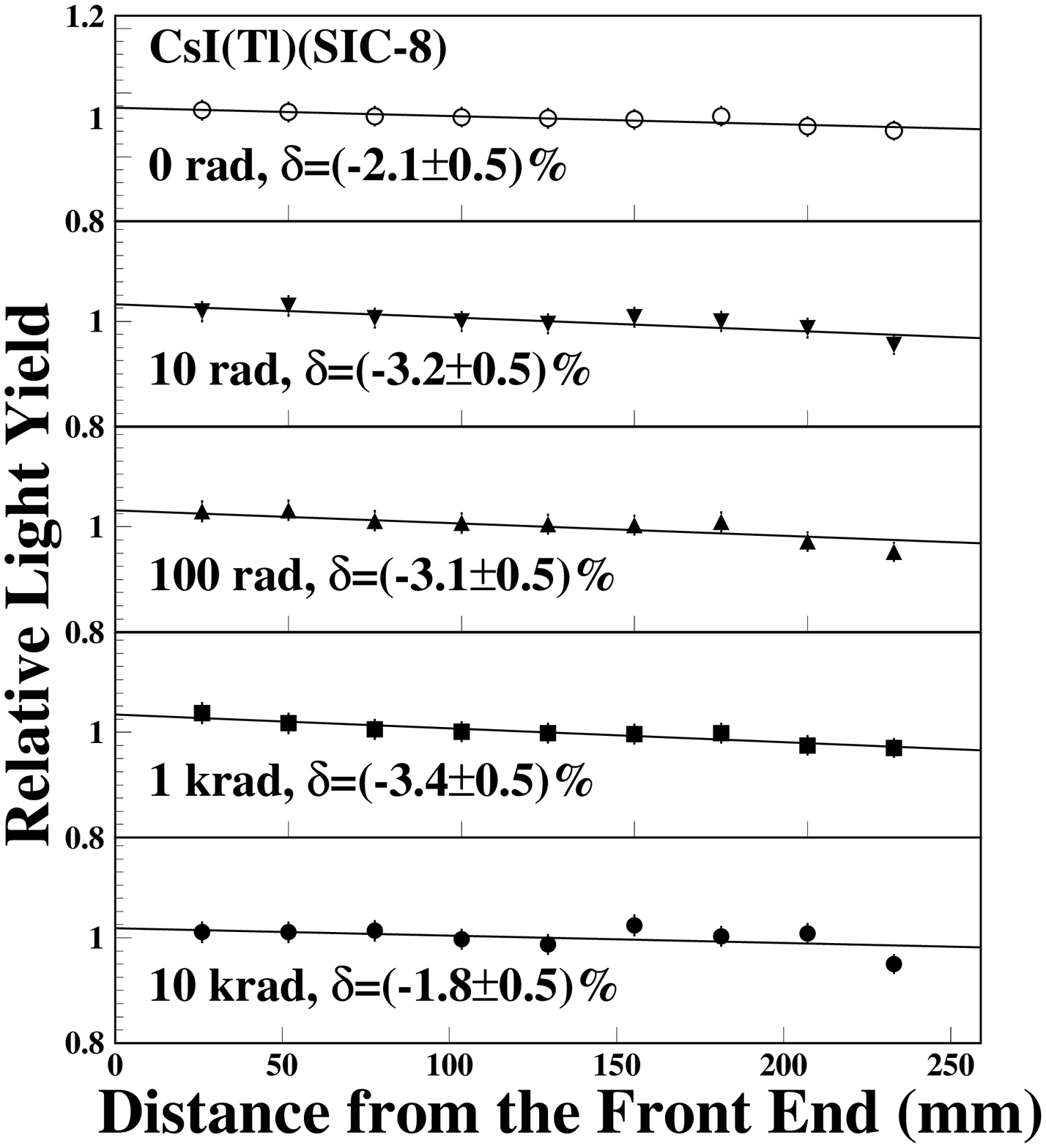}
\includegraphics{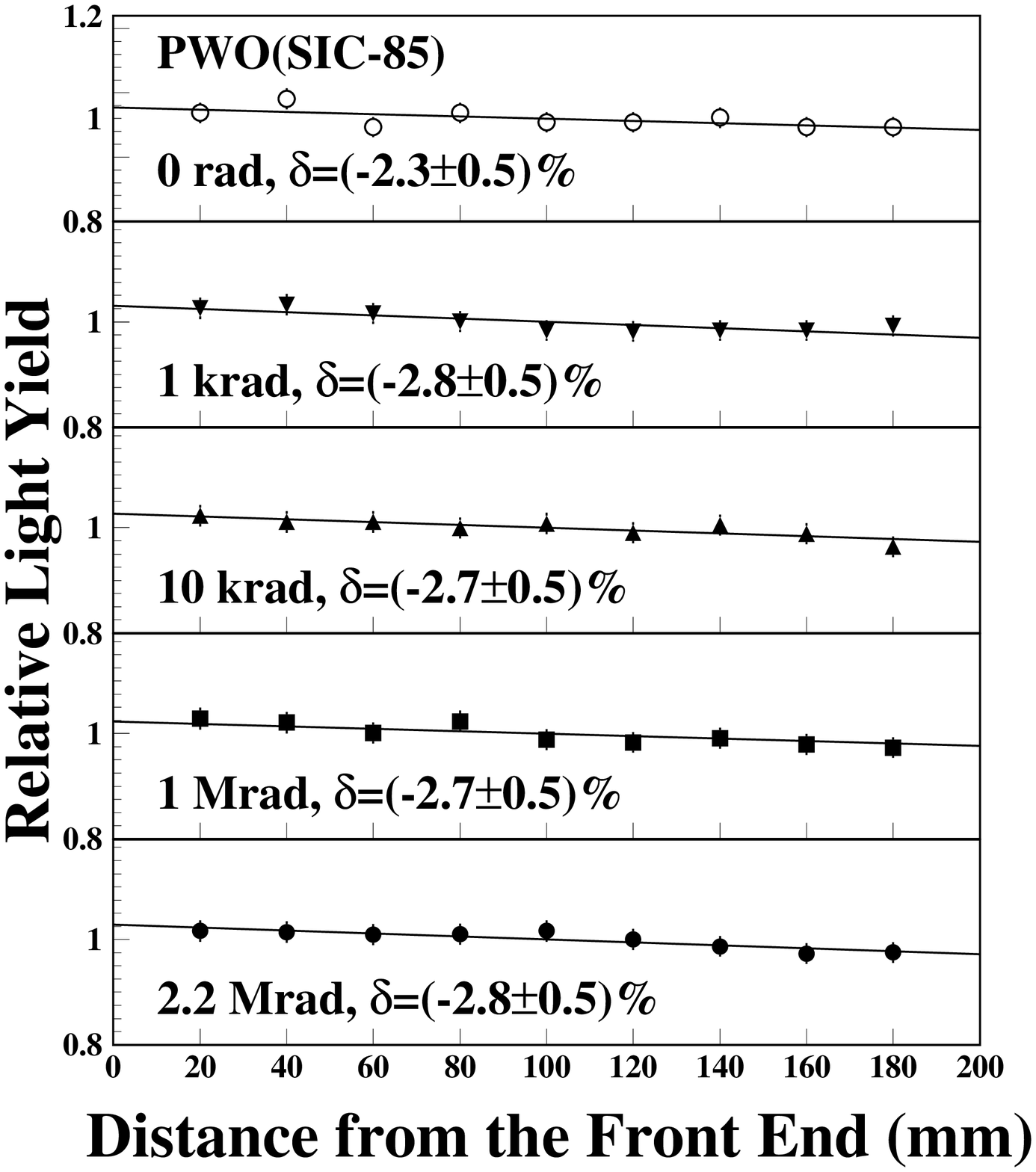}
\caption{Light response uniformities are shown as a function 
 of integrated dose for full size CsI(Tl)~\protect\cite{bib:zhu_csi(tl)}
 (left) and \PBW~\protect\cite{bib:zhu_pwo} (right) samples.}
\label{fig:uniformity_rad}
\end{figure}

First, scintillation mechanism in scintillating crystals is usually
not damaged by radiation. Degradation of light output is thus 
due only to radiation-induced absorption, i.e. color 
center formation. As a consequence, irradiation does not change light 
response uniformity. Figure~\ref{fig:uniformity_rad} shows light response 
uniformity as a function of accumulated dose for full size CsI(Tl) (left)
and \PBW (right) crystals. Pulse heights measured in nine points 
evenly distributed along the longitudinal axis of the crystal is fit to 
Equation~\ref{eq:unit}, showing clearly that the slope ($\delta$) does not 
change up to 10 krad for a CsI(Tl) sample, even only the front few cm of 
the sample was irradiated~\cite{bib:zhu_csi(tl)}, and to 2.2 Mrad for 
a \PBW sample~\cite{bib:zhu_pwo}. This result is understood, as the 
intensity of all light rays attenuates equally after 
passing the same radiation-induced absorption zone in the crystal. 
A ray-tracing simulation shows that the slope of light response uniformity 
depends only on crystal geometry for crystals with long enough light 
attenuation length, and will change only if light attenuation 
length degrades to less than about 4 times crystal length~\cite{bib:zhu_nim}. 
This leads to a conclusion that crystal's energy resolution would not 
degrade by radiation although its calibration does change, which was 
confirmed by beam test at CERN~\cite{bib:ecal_tb}. 
Since degradation of the amplitude of light output can 
be inter-calibrated with physics events, or by a light monitoring system
if it is caused by optical absorption, crystal precision can be maintained 
{\it in situ} even radiation damage does occur.

Second, the level of light output degradation under continuous 
irradiation of certain dose rate approaches an equilibrium, 
leading to a dose rate dependent damage, which was also later 
confirmed in CERN beam test~\cite{bib:seez_tb}. 
This ``dose rate dependence'' of light output degradation is understood 
to be caused by color center kinetics of the creation and annihilation of
radiation induced color centers~\cite{bib:zhu_pwo,bib:zhu_nim}.

\ \\

If both annihilation and creation coexist, the color center density 
at the equilibrium depends on the dose rate applied. Assuming  
annihilation speed of a color center $i$ is proportional to a constant 
$a_i$ and its creation speed is proportional to a constant $b_i$ 
and dose rate ($R$), the differential change of color center 
density when both processes coexist can be written as \cite{bib:zhu_bleach}:
\begin{equation}
d D = \sum_{i=1}^{n} \{- a_i  D_i  + (D^{all}_i - D_i)~ b_i R \}dt,
\label{eq:density}
\end{equation}
where $D_i$ is the density of the color center $i$ in the crystal
and the summation goes through all centers. The solution of 
Equation~\ref{eq:density} is
\begin{equation}
D =  \sum_{i=1}^{n} \{\frac{b_i R D^{all}_i}{a_i + b_i R}~[ 1 
 - e^{-(a_i  + b_i R) t}] + D^0_i e^{-(a_i + b_i R) t}\},
\label{eq:solution}
\end{equation}
where $D^{all}_i$ is the total density of the trap related to the 
center $i$ and $D_i^0$ is its initial density. The color center density 
in equilibrium ($D_{eq}$) thus depends on the dose rate ($R$).
\begin{equation}
 D_{eq} = \sum_{i=1}^n \frac{b_i R D^{all}_i}{a_i + b_i R}, 
\label{eq:kinetics}
\end{equation}
By using color center kinetics, one can calculate, or predict, 
crystal damage at one dose rate by using data collected at
another dose rate~\cite{bib:pwo_dengq}. 

\section{Damage Mechanism in Scintillating Crystals}
\label{sec:mechanism}

Understanding damage mechanism in scintillators
would help to improve quality of mass produced crystals, which is
usually achieved by material analysis. 
Glow Discharge Mass Spectroscopy (GDMS) analysis was tried in Charles 
Evans \& Associates and Shiva Technology, looking for correlations 
between the trace impurities in crystals and their radiation hardness. 
Samples were taken 3 to 5 mm below the surface of the crystal to avoid 
surface contamination. For both CsI(Tl) and \PBW crystals, a survey of 76 
elements, including all of the lanthanides, indicates that there are no 
obvious correlations between the detected trace impurities and crystal's 
susceptibility to the radiation damage. This indicates possible
role of other defects, such as oxygen contamination or stoichiometric
vacancies, which can not be  determined by GDMS.

\subsection{Damage Mechanism in Alkali Halides}

Oxygen contamination is known to cause radiation damage 
in alkali halide scintillators. In \BAF~\cite{bib:zhu_baf}, for example, 
hydroxyl (OH$^-$) may be introduced into crystal through a hydrolysis 
process, and latter decomposed to interstitial and substitutional centers 
by radiation through a radiolysis process. Equation~\ref{eq:baf} shows
a scenario of this process: 
\begin{equation}
 OH^{-} \rightarrow  H_{i}^{0} + O_s^{-}~~or~~H_{s}^{-} + O_{i}^{0},
\label{eq:baf}
\end{equation}
where subscript $i$ and $s$ refer to interstitial and substitutional 
centers respectively. Both $O_s^{-}$ and U ($H_{s}^{-}$) centers were 
identified~\cite{bib:zhu_baf}.  

\begin{figure}
\vspace{9.5cm}
\includegraphics{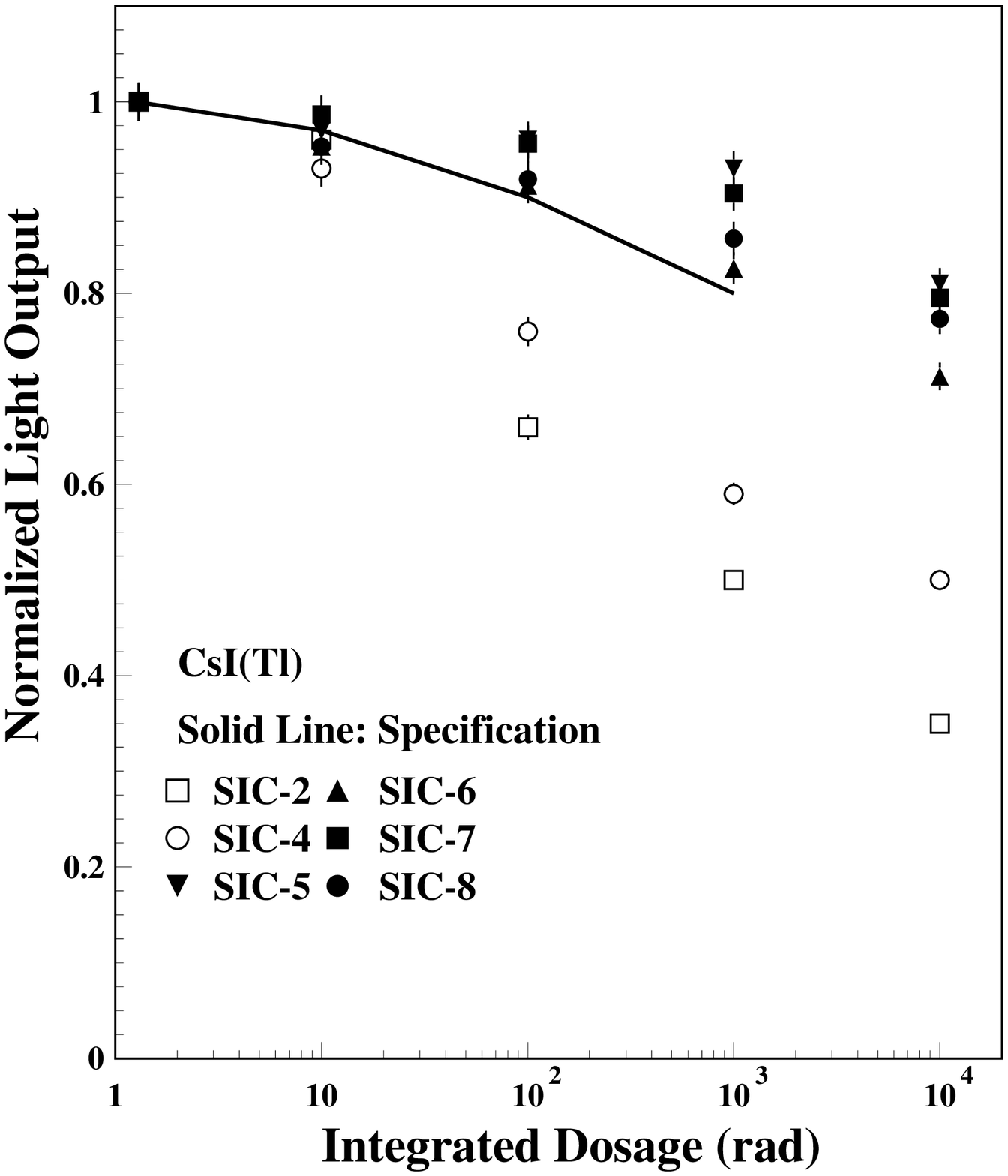}
\includegraphics{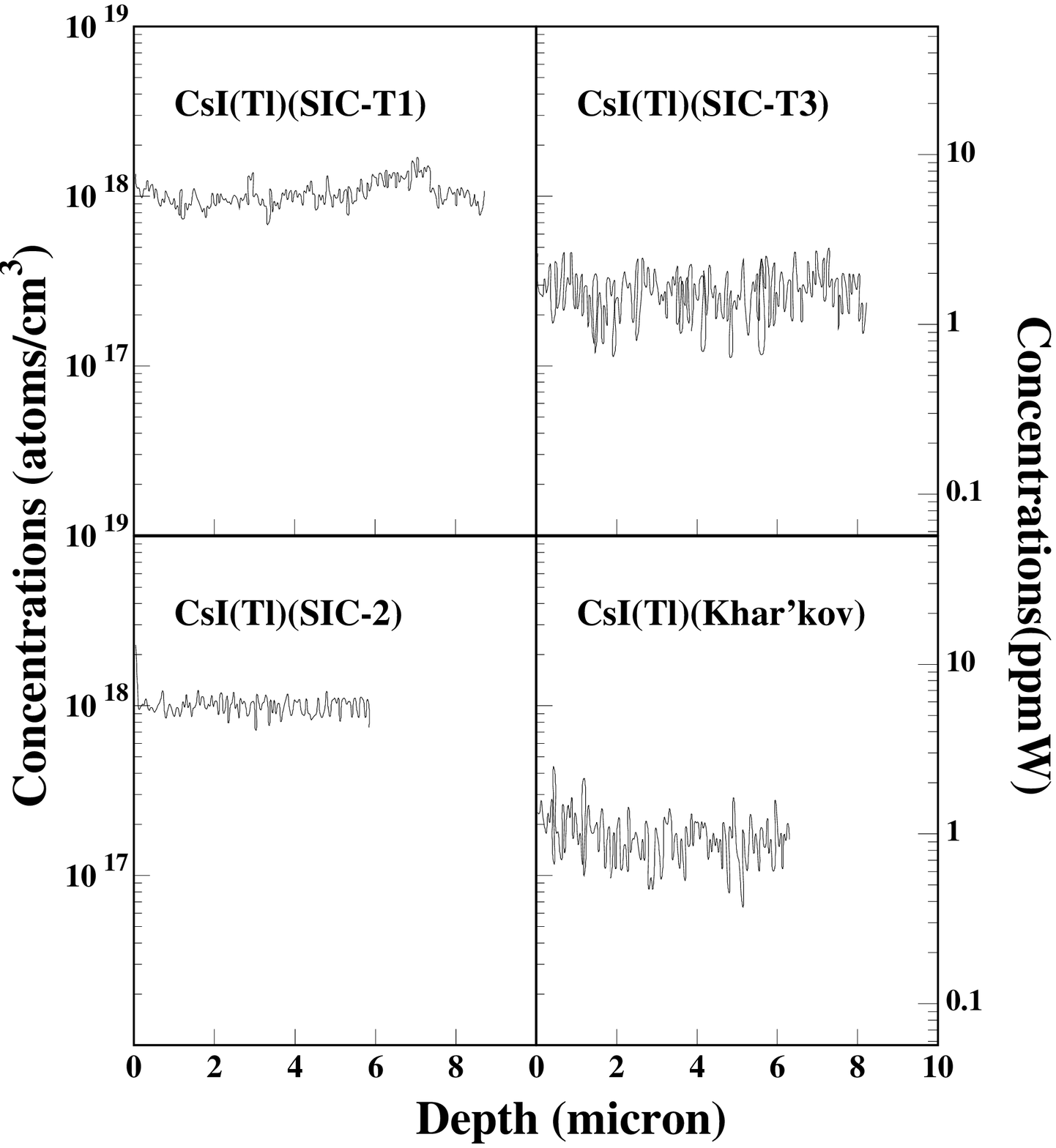}
\caption{Left: The progress of CsI(Tl) radiation hardness is shown for 
 full size ($\sim$30 cm) CsI(Tl) samples from SIC together with the rad-hard 
 specification of $BaBar$ experiment. Right:
 Right: The depth profiles of oxygen in CsI(Tl) samples measured
 at Charles Evans \& Associates by using SIMS analysis.}
\label{fig:csi}
\end{figure}

Following \BAF experience, effort was made to remove oxygen contamination 
in CsI(Tl) crystals. A scavenger was used at SIC to remove oxygen contamination,
leading to significant improvement of CsI(Tl) quality~\cite{bib:zhu_csi(tl)}.  
The left side of Figure~\ref{fig:csi} shows the light output as 
a function of accumulated dose for full size CsI(Tl) samples, compared 
to the $BaBar$ radiation hardness specification (solid line). While the late 
samples SIC-5, 6, 7 and 8 (with scavenger) satisfy the $BaBar$ 
specification, early samples SIC-2 and 4 did not. The function of the 
scavenger is to form  oxide with density less than CsI, so will migrate 
to the top of ingot during growing process, similar to zone-refining. 
By doing so, both oxygen and scavenger are removed from the crystal.

Quantitative identification of oxygen contamination in CsI(Tl) samples
needs additional analysis. Gas Fusion (LECO) at Shiva Technologies West, 
Inc., found that oxygen contamination in all
CsI(Tl) samples is below detection limit of 50 ppm.
Secondary Ionization Mass Spectroscopy (SIMS) was tried at
Charles Evans \& Associates.  A Cs ion beam of 6 keV and 50 nA was 
used to bombard the CsI(Tl) sample. All samples were freshly cleaved 
prior before being loaded to the UHV chamber. An area of 
0.15 $\times$ 0.15 mm$^2$ on the cleaved surface was analyzed.
To further avoid surface contamination, the starting point of 
the analysis is at about 10 $\mu$m deep inside the fresh cleaved surface. 
The right side of Figure~\ref{fig:csi} shows depth profile of
oxygen contamination for two rad-soft (SIC-T1 and SIC-2) and two rad-hard 
(SIC-T3 and Khar'kov) CsI(Tl) samples. Crystals with poor radiation 
resistance have oxygen contamination of 10$^{18}$ atoms/cm$^3$ or 
5.7 ppmW, which is 5 times higher than the  background count 
(2$\times10^{17}$ atoms/cm$^3$, or 1.4 ppmw).  The radiation damage 
in CsI(Tl) is indeed caused by oxygen contamination. 

\subsection{Damage Mechanism in Oxides}

Crystal defects, such as oxygen vacancies, is known to cause radiation 
damage in oxide scintillators. In BGO, for example,  three common 
radiation induced absorption bands  at 2.3, 3.0 and 3.8 eV  were 
found in a series of 24 doped samples \cite{bib:zhu_bgo},
indicating  defect-related color centers, such as oxygen vacancies. 
Following the BGO experience, an effort was made at SIC to reduce oxygen 
vacancies in \PBW crystals by oxygen compensation through post-growth 
thermal annealing in an oxygen-rich atmosphere, and result was 
positive~\cite{bib:zhu_pwo}. 

Particle Induced X-ray Emission (PIXE) and quantitative 
wavelength  dispersive Electron Micro-Probe Analysis (EMPA) 
was tried in Charles Evans \& Associates to quantify
stoichiometry deviation and oxygen vacancies in \PBW crystals. 
Crystals with poor radiation hardness were indeed found to have a 
non-stoichiometric W/Pb ratio~\cite{bib:zhu_pwo}. However, both 
PIXE and EMPA did not provide oxygen analysis. X-ray Photoelectron 
Spectroscopy (XPS) at Charles Evens \& Associates was found to be 
very difficult to reach a stable quantitative conclusion because of 
large systematic uncertainties in oxygen analysis~\cite{bib:xps}. 

\begin{figure}[ht]
\ \\
\includegraphics*[bbllx=130pt,bblly=210pt,bburx=470pt,bbury=600pt,
 width=0.48\textwidth]{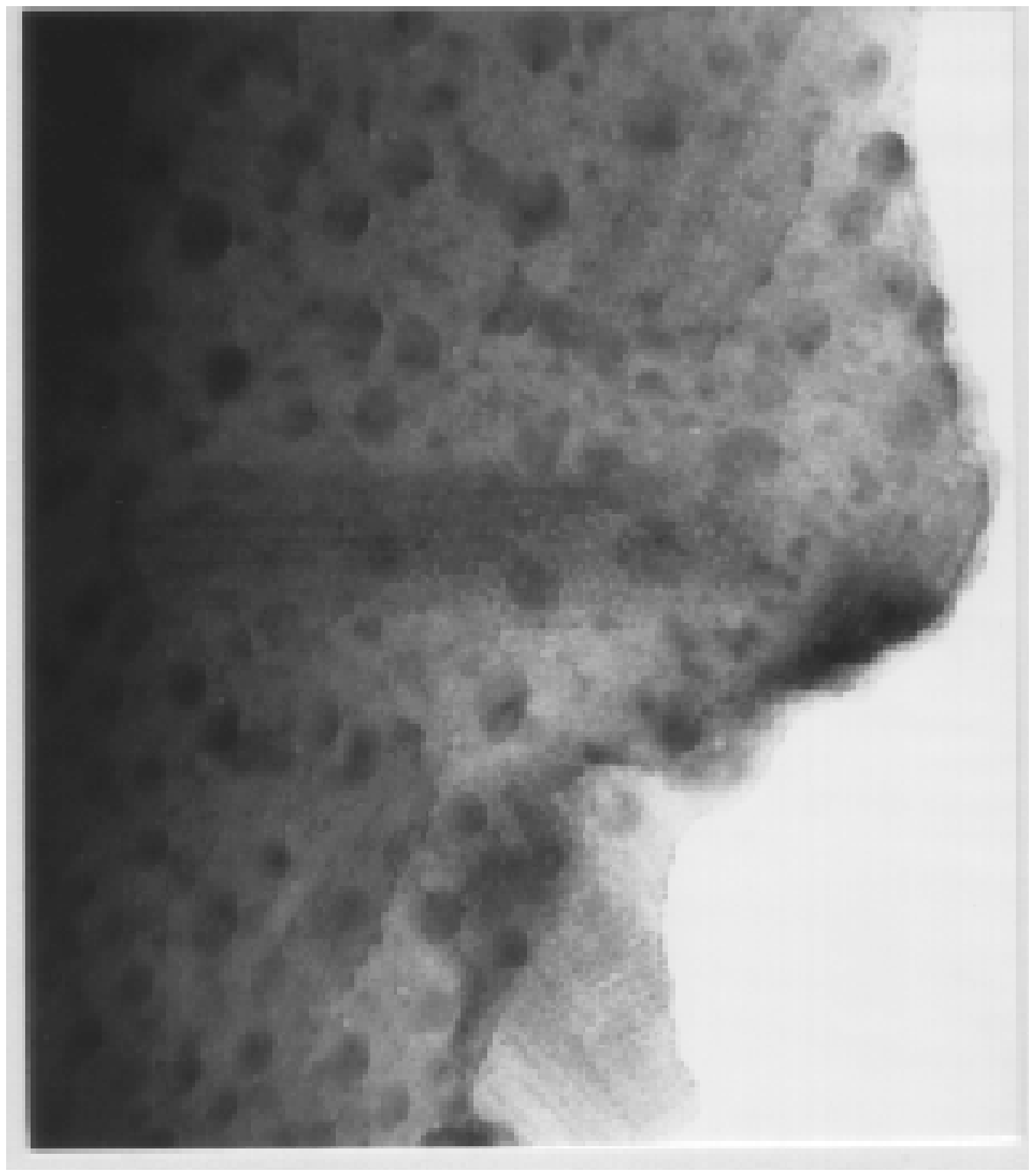}
\includegraphics*[bbllx=130pt,bblly=210pt,bburx=470pt,bbury=600pt,
 width=0.48\textwidth]{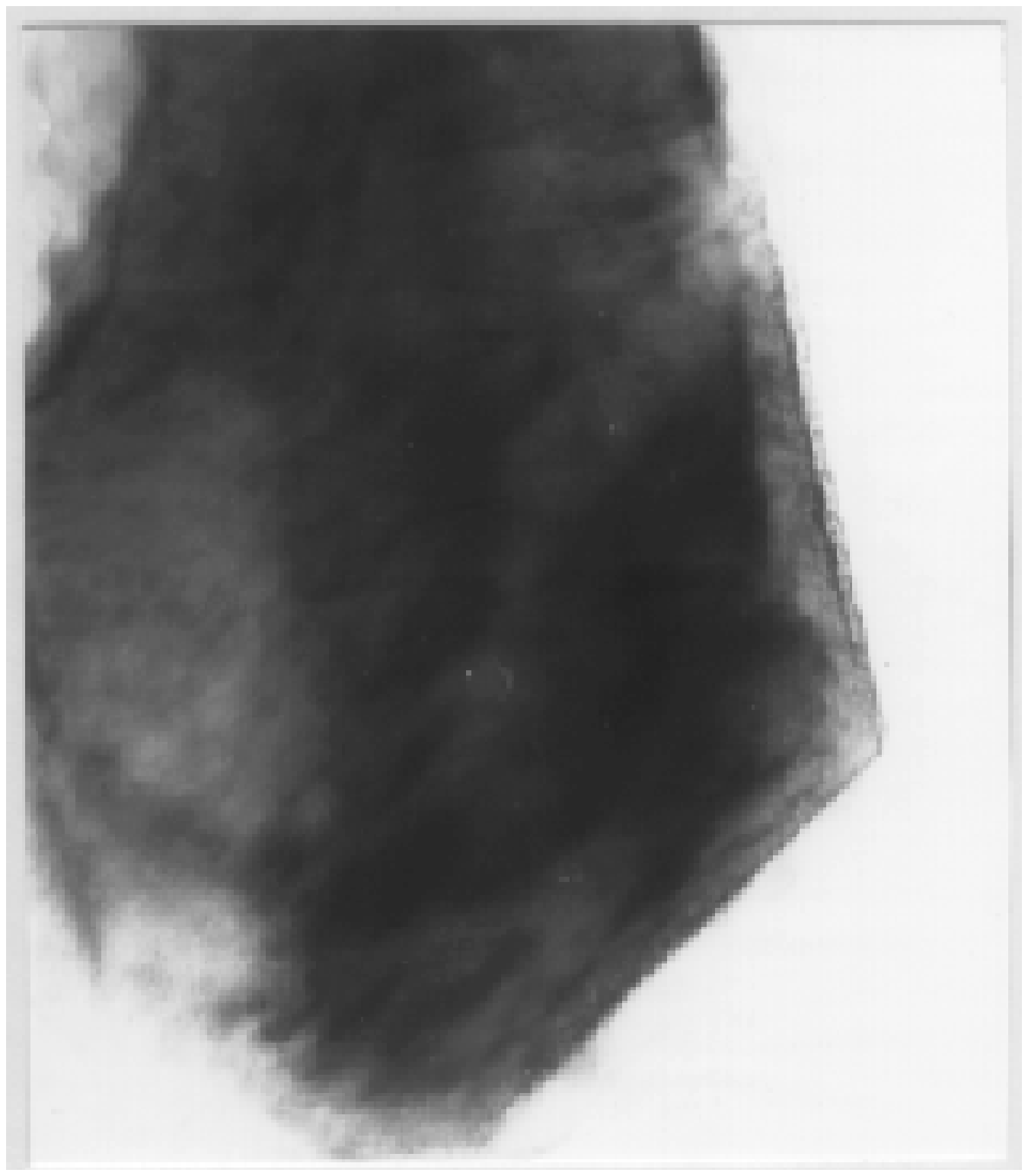}
\ \\
\caption{TEM pictures of a \PBW crystal of poor (left) 
 radiation hardness, showing clearly the black spots of $\phi$5--10 nm 
 related to  oxygen vacancies, as compared to that of a good one (right).} 
\label{fig:tem}
\end{figure}

By using Transmission Electron Microscopy (TEM) a localized  stoichiometry 
analysis was possible to identify oxygen vacancies. A TOPCON-002B Scope was 
first used at 200 kV and 10 $\mu$A. Samples were made to powders 
of an average grain size of a few $\mu$m, and then placed on a sustaining
membrane. Figure~\ref{fig:tem} shows TEM pictures taken for a pair of samples 
of poor (left) and good (right) radiation hardness. Black spots of a diameter 
of 5 -- 10 nm were clearly observed in the poor sample, but not in the 
good sample. These black spots were identified as regions with severe 
oxygen deficit by a localized stoichiometry analysis using TEM coupled 
to Energy Dispersion Spectrometry (EDS)\cite{bib:sic_tem}. Approaches 
to reduce oxygen deficits were taken by crystal vendors, leading to 
production crystals of much improved quality. 

\section{Summary}
\label{sec:summary}

Precision crystal calorimetry extends physics reach 
in experimental high energy physics because of 
its best achievable resolution for electrons and photons.
An optimized light response uniformity is the key to reach
crystal energy resolution. A precision calibration is 
the key to maintain crystal  precision {\it in situ}.

Predominant radiation damage 
effect in crystal scintillators is radiation induced absorption, 
or color center formation, not damage of scintillation mechanism. 
For precision calorimetry, crystal scintillator must preserve its
light response uniformity under irradiation, which requires a long 
enough initial light attenuation length and a low enough radiation
induced color center density. A precision light monitoring may function 
as inter-calibration for such crystals.

Radiation damage in alkali halides is caused 
by oxygen and/or hydroxyl contamination,  as evidenced by a SIMS 
analysis and the effectiveness  of a scavenger in removing 
oxygen contamination in CsI(Tl) crystals. Radiation damage in oxides 
is caused by stoichiometry-related defects, e.g.  
oxygen vacancies,  as evidenced by a localized stoichiometry analysis 
using TEM/EDS, and the effectiveness of the oxygen compensation for 
\PBW crystals.

\section*{Acknowledgements}

Measurements at Caltech were carried out
by Mr. Q.~Deng, H~Wu, D.A.~Ma, Z.Y.~Wei and T.Q.~Zhou.
Part of the \PBW related work was carried out by Dr. C. Woody 
and his group at Brookhaven National Laboratory.

\end{document}